\documentclass[onecolumn,showpacs,amsfonts,aps,prc,nofootinbib,floatfix,%
superscriptaddress]{revtex4}
\usepackage{epsfig}
\usepackage{amsmath}
\usepackage{bm}
\usepackage{graphicx}

\newcommand{\beq}{\begin{equation}}
\newcommand{\eeq}{\end{equation}}
\newcommand{\bea}{\vspace{0.25cm}\begin{eqnarray}}
\newcommand{\eea}{\end{eqnarray}}

\newcommand{\ro}{\mbox{{\boldmath
$\rho$}}}

\newcommand{\qb}{\mbox{{\bf
q}}}
\newcommand{\pb}{{{\bf p}}}

\def\lsim{\mathrel{\rlap{\lower4pt\hbox{\hskip1pt$\sim$}}
    \raise1pt\hbox{$<$}}}         
\def\gsim{\mathrel{\rlap{\lower4pt\hbox{\hskip1pt$\sim$}}
    \raise1pt\hbox{$>$}}}         

\long\def\symbolfootnote[#1]#2{\begingroup%
\def\thefootnote{\fnsymbol{footnote}}\footnotemark[#1]\footnotetext[#1]{#2}\endgroup}

\begin{document}

\title{
Jet quenching for heavy flavors
in $AA$ and $pp$ collisions
}

\author{B.G. Zakharov}

\address{
L.D.~Landau Institute for Theoretical Physics,
        GSP-1, 117940,\\ Kosygina Str. 2, 117334 Moscow, Russia
}

\begin{abstract}
We perform a global analysis of experimental
data on jet quenching for heavy flavors
for scenarios with
and without quark-gluon plasma formation in $pp$ collisions.
We find that the theoretical predictions for the nuclear modification
factor $R_{AA}$ for heavy flavors at the LHC energies
are very similar for these scenarios, and
the results for $R_{AA}$ and $v_2$ agree reasonably
with the LHC data. The agreement with data at the RHIC top energy
becomes somewhat better for the
intermediate scenario, in which the quark-gluon plasma formation in
$pp$ collisions occurs only at the LHC energies.
Our fits to heavy flavor $R_{AA}$
show that description of jet quenching for heavy flavors requires
somewhat bigger $\alpha_s$ than data on jet quenching for light hadrons.

\end{abstract}
%

\maketitle
\section{Introduction}
The observed suppression of high-$p_T$ hadron spectra
(jet quenching) in nucleus-nucleus ($AA$) collisions at RHIC and the LHC 
is one of the main signals of formation of a deconfined quark-gluon
plasma (QGP) in the initial stage of $AA$ collisions.
Jet quenching in $AA$ collisions is due to
radiative \cite{BDMPS1,LCPI1,W1,GLV1,AMY1,BSZ} and collisional
\cite{Bjorken1} energy loss of fast partons
traversing the QGP fireball. The dominant contribution to
the parton energy loss comes from induced gluon radiation
\cite{BSZ,Z_Ecoll}.
The suppression of particle spectra in $AA$ collisions
as compared to the binary scaled spectra in $pp$ collisions
is characterized by the
nuclear modification factor $R_{AA}$.
Experimentally, for a centrality class
$\Delta c$, $R_{AA}$ is defined as
\beq
R_{AA}=\frac{d^2N_{AA}/dp_T^2dy}{N_{ev}
  \langle T_{AA}\rangle_{\Delta c} d^2\sigma_{pp}/dp_T^2dy},
\label{eq:10}
\eeq
where $N_{ev}$ is the number of events, $d^2N_{AA}/dp_T^2dy$ is
the particle yield in $AA$ collisions,
$\langle T_{AA}\rangle_{\Delta c}$ is the averaged (over the centrality class
$\Delta c$) nuclear overlap function. The centrality $c$, which
characterizes the overlap of the colliding nuclei, is experimentally
determined via charged hadron multiplicity.
For heavy ion collisions,
to good accuracy the centrality can be written via the impact parameter $b$
as $c\approx \pi b^2/\sigma_{in}^{AA}$ \cite{centrality}
(except for very peripheral collisions).
If one assumes that in proton-proton ($pp$) collision the QGP
is not produced, and the experimental inclusive $pp$ cross section
in the denominator of (\ref{eq:10}) is close to the
inclusive $pp$ cross
calculated within the pQCD framework,
$d^2\sigma_{pp}^{pt}/dp_T^2dy$,
then the theoretical nuclear modification factor
can be written as
\beq
R_{AA}=\frac{\langle d^2\sigma_{NN}^{m}/dp_T^2dy\rangle_{\Delta c}}
{d^2\sigma_{pp}^{pt}/dp_T^2dy},
\label{eq:20}
\eeq
where
$d^2\sigma_{NN}^{m}/dp_T^2dy$ is the medium-modified inclusive nucleon-nucleon cross section
for a given geometry of the jet production in $AA$ collision, and 
$\langle\dots\rangle$ means averaging over the jet production
geometry and the impact parameter for the centrality bin $\Delta c$.

If the QGP formation occurs in $pp$
collisions as well, formula (\ref{eq:20}) becomes invalid, since in this
scenario the $pp$ cross section in the denominator of (\ref{eq:10})
is affected by the medium effects, and one should use in the
denominator of (\ref{eq:20}) instead of the pQCD $pp$ cross section the one
that accounts for
jet modification by the final state interaction medium effects
in the mini QGP (mQGP).
Several signals of the mQGP formation
in $pp$ collisions have by now been seen in data on soft hadron
production. Among them the observation of the ridge effect
\cite{CMS_ridge,ATLAS_mbias} in $pp$ collisions at the LHC energies,
the steep growth  of the strange particle production
at $dN_{ch}/d\eta\sim 5$ \cite{ALICE_strange}.
The latter fact agrees with the onset of the QGP regime
at $dN_{ch}/d\eta \sim 6$   predicted in \cite{Camp1}
from experimental data  on
the mean $p_{T}$ as a function of multiplicity,
employing Van Hove's arguments \cite{VH}.
From the point of view of the mQGP formation, it is important that
in the $pp$ jet events multiplicity of soft (underlying-event (UE)) hadrons
is bigger than multiplicity in minimum bias $pp$ collisions by a factor of $\sim 2-2.5$
\cite{Field}. At the LHC energies $dN_{ch}^{ue}/d\eta\sim 10-15$,
which turns out to be well above the estimated critical multiplicity density
$dN_{ch}/d\eta\sim 5$ for the onset
of the mQGP formation in $pp$ collisions.
For $pp$ collisions at the RHIC top energy of $\sqrt{s}=0.2$ TeV we
have $dN_{ch}^{ue}/d\eta\sim 6$, which is of the order of the expected
multiplicity for the onset of the QGP formation regime.
Thus, it is possible that for $pp$ collisions at $\sqrt{s}\sim 0.2$ TeV
the dynamics of the produced soft hadrons may be close to the free
streaming regime, and consequently the jet quenching effects should be small.
This means that for $AA$ collisions at RHIC the
theoretical $R_{AA}$ should be given by the formula (\ref{eq:20}).

For the scenario with the mQGP production in $pp$ collisions,
the real inclusive $pp$ cross section in the denominator of (\ref{eq:10})
includes the jet quenching effects in the mQGP fireball produced in
$pp$ collision. 
We can write it as the product of the theoretical pQCD $pp$ cross
section and the medium modification factor $R_{pp}$
\beq
d^2\sigma_{pp}^m/dp_T^2dy=R_{pp}
{d^2\sigma_{pp}^{pt}/dp_T^2dy}.
\label{eq:30}
\eeq
Physically, the $d^2\sigma_{pp}^m/dp_T^2dy$ is similar to the effective $NN$
cross section entering the nominator of (\ref{eq:20}), but, contrary to
(\ref{eq:20}), now we should perform calculations for the mQGP fireball and
perform averaging over the geometry of $pp$ collisions.  Thus, in the
scenario with the mQGP production in $pp$ collisions the theoretical $R_{AA}$,
as compared to the formula
(\ref{eq:20}), turns out to be
enhanced by the factor $1/R_{pp}$.
Of course, $R_{pp}$ is not directly observable quantity.
Since the size and the temperature of the mQGP fireball in $pp$ collisions
should be small, one can expect that the quenching effects should be small,
i.e. $R_{pp}$ should be close to unity.
This makes it practically impossible the observation of jet quenching in $pp$
collisions via experimental data on the $p_T$-dependence of hadron spectra.
In \cite{Z_pp_PRL} it was shown that measurement of 
variation of the photon/hadron-tagged jet
fragmentation functions (FFs), characterized by the medium modification
factor $I_{pp}$, with the UE multiplicity
may be a promising method for direct observation of the jet quenching
in $pp$ collisions.
Recently, the ALICE Collaboration  reported preliminary results \cite{ALICE_Ipp}
on the medium modification factor $I_{pp}$ at $\sqrt{s}=5.02$ TeV
for the hadron-tagged jets (with the trigger hadron momentum $8<p_T<15$ GeV,
and the associated away side hadron momentum in the range  $4<p_T<6$ GeV),
that show a monotonic decrease of $I_{pp}$ with the UE multiplicity
by about 15-20\% for the UE multiplicity density range $\sim 4-15$.
In \cite{Z_Ipp} it has been shown that this agrees reasonably
with theoretical predictions obtained within the light cone path integral
(LCPI) \cite{LCPI1} approach to induced gluon emission.
The observation of the decrease of $I_{pp}$ with the UE multiplicity,
if confirmed, will be a strong argument for the scenario with the mQGP
production in $pp$ jet events. 

In the light of the possibility of the mQGP formation in $pp$ collisions,
it is of great interest to perform analysis of jet quenching in $AA$
collisions for such a scenario. In \cite{Z_hl}, we have performed
the global analysis of the
data on jet quenching in $AA$ collisions for light hadrons for
scenarios with and without the mQGP production in $pp$ collisions 
within the LCPI  approach \cite{LCPI1} to induced
gluon emission.
We used $\alpha_s(Q,T)$ which has a plateau
around $Q\sim Q_{fr}= \kappa T$
(this form is motivated by the lattice results
for the in-medium QCD coupling \cite{Bazavov_al1} and calculations
within the functional renormalization group \cite{RG1}).
We fitted $\kappa$ using
the LHC heavy ion data on the nuclear modification factor $R_{AA}$
in $2.76$ and $5.02$ TeV Pb+Pb, and $5.44$ TeV Xe+Xe collisions.
Calculations in this way allow to avoid the
ambiguities in the choice of $\alpha_s$ for small systems,
because the parameter $\kappa$, fitted to data for heavy ion collisions,
automatically fixes $\alpha_s$ for small size QGP.
In \cite{Z_hl} it was found that both the models lead to quite good
description of the RHIC and the LHC data on $R_{AA}$ for heavy ion collisions.
For the RHIC PHENIX data on $R_{AA}$ the agreement becomes somewhat better
for a scenario when
the mQGP formation in $pp$
collisions occurs at the LHC energies, but is absent for the RHIC energies.

It would be interesting to examine whether 
the scenario with the mQGP formation in $pp$ collisions
is consistent with the data on jet quenching for heavy flavors as well.
Jet quenching for heavy flavors has attracted much theoretical and
experimental attention in recent years
(for recent review, see \cite{Apolin_HQ}). Initially it was expected
 that heavy quarks should lose less
 energy than light quarks due to the dead cone suppression of the
 radiative energy loss for heavy quarks \cite{DK}.
However, later experiments at  RHIC \cite{PHENIX1_e,STAR_e} 
 observed a quite strong suppression  
 of single electrons from decays of heavy mesons 
 that seemed to be in contradiction 
with expected dead cone suppression  
of the radiative energy loss (the ``heavy quark puzzle'').
On the theoretical side,
in \cite{AZ} within the LCPI approach \cite{LCPI1} to the induced gluon emission
it was found that, due to the quantum finite-size effects
(ignored in the dead cone model \cite{DK}), at low energies ($\lsim 20-30$
GeV) the quark mass suppression of radiative energy loss 
 turns out to be significantly smaller than 
predicted in the dead cone model. Moreover, at energies $\gsim 100$ GeV
the quantum effects lead to an increase of the radiative energy loss with
the quark mass.
In Refs. \cite{RAA12,RAA13} we analyzed the first data on jet quenching for
heavy flavors from the LHC within the LCPI approach for the scenario
without the mQGP production in $pp$ collisions, and found
a reasonable agreement with the data. To date, a substantial amount
of experimental data on jet quenching for heavy flavors has been obtained
at the LHC. This allows to perform a
more detailed comparison of theory and experiment for the heavy flavor
jet quenching.
In the context of the heavy quark puzzle, it is important that
the scenario with the mQGP formation can lead to some reduction of the 
heavy-to-light ratios of the nuclear modification factors $R_{AA}$
\cite{hq16}.
This occurs due to the flavor hierarchy 
$R_{pp}^\pi<R_{pp}^D<R_{pp}^B$ \cite{hq16}, which is valid at $p_T\lsim 20$ GeV
for the RHIC energy $\sqrt{s}=0.2$ TeV and at $p_T\lsim 70$ GeV for the LHC
energies \cite{hq16}.

In this paper we extend the analysis of \cite{Z_hl}
of jet quenching for light hadrons to heavy mesons
and heavy flavor electrons (HFEs).
As in \cite{Z_hl}, we calculate the induced gluon emission $x$-spectrum, $dP/dx$
($x$ is the gluon fractional momentum), within the LCPI approach \cite{LCPI1}
(see also \cite{Z2019} for a more recent discussion
of the LCPI formalism). In this approach $dP/dx$ is expressed through the
solution of a two-dimensional Schr\"odinger equation,
which automatically accounts for all rescatterings of fast partons
in the medium.
We calculate the induced gluon
spectrum using the form suggested in \cite{Z04_RAA}\footnote{
  Contrary to the original LCPI  form of the induced gluon
  spectrum in terms of the singular Green functions \cite{LCPI1}, the method
  of \cite{Z04_RAA} reduces calculation of the gluon spectrum
  to solving an initial boundary value problem with a smooth initial
  condition, which is convenient for numerical calculations.}.
We calculate the induced gluon spectrum beyond the soft gluon approximation.
In the literature the heavy quark energy loss is usually calculated in the
soft gluon approximation (see e.g.
\cite{CUJET3,MDjordj1,Blok1,Blok2,Blok3,Rapp_HQ,Vitev_SCET}). 
However, one can easily show that this approximation
is too crude for analysis of the quark mass effects. 
Indeed, in the two-dimensional Schr\"odinger equation,
which defines the induced gluon $x$-spectrum,
the quark mass enters only through the formation length
$L_{f}=2x(1-x)E/[m_{q}^{2}x^{2}+m_{g}^{2}(1-x)]$ \cite{LCPI1} (here
$E$ is the initial quark energy, 
$m_{q,g}$ are the quasiparticle parton masses).
For this reason, the quark mass becomes important at 
$x^{2}/(1-x)\gsim m_{g}^{2}/m_q^{2}$. 
Taking $m_{g}\sim 400$ MeV \cite{LH}, one can see 
that for $c(b)$-quark it occurs 
at $x\gsim 0.3(0.1)$  (accurate computations of \cite{AZ} corroborate
these qualitative estimates).
This says that the soft gluon approximation
may be unsatisfactory for heavy flavors (especially for $c$-quark). 
Note also that our scheme treats accurately the Coulomb effects
in parton rescatterings (contrary to available in
the literature \cite{Blok1,Blok2,Blok3}  perturbative treatment of the
Coulomb effects as a correction to the harmonic oscillator approximation),
that are very important for the quark mass effects \cite{AZ}.

The plan of the paper is as follows.  In section 2, we briefly review
the basic aspects of our model.
In section 3
we present results for $R_{pp}$ and comparison of our results
with experimental data on $R_{AA}$ and on
the elliptic flow coefficient $v_2$ in $AA$ collisions.
Section 4 presents a summary.

\section{Outline of the jet quenching model}
We use the jet quenching scheme of \cite{RAA08} in the
form of \cite{RAA20} with a somewhat improved treatment of multiple
gluon emission and adopted for use of a $T$-dependent $\alpha_s$
(as in \cite{Z_hl}).
In this section we briefly discuss the basic features of our theoretical
scheme. More details can be found in Refs. \cite{RAA08,RAA20,Z_hl}.

For a given geometry of the $AA$ collision and of the jet production
we write the medium-modified hard cross section for $NN$ collision 
in a form similar to the ordinary pQCD formula for $NN$ collisions in vacuum
\beq
\frac{d\sigma^{m}(N+N\rightarrow h+X)}{d\pb_{T} dy}=
\sum_{i}\int_{0}^{1} \frac{dz}{z^{2}}
D_{h/i}^{m}(z, Q)
\frac{d\sigma^{pt}(N+N\rightarrow i+X)}{d\pb_{T}^{i} dy}\,,\,\,\,
\label{eq:40}
\eeq
where
${d\sigma^{pt}(N+N\rightarrow i+X)}/{d\pb_{T}^{i} dy}$ is the standard pQCD
hard cross section for production of the initial hard parton $i$ with the
transverse
momentum $\pb_{T}^{i}=\pb_{T}/z$,
$D_{h/i}^{m}$ is the medium-modified
FF describing
the production of the observed particle $h$ from the fragmentation
of the initial hard parton $i$.
For the initial virtuality scale $Q$ we use the parton momentum $p^{i}_{T}$.
We calculate hard cross sections using the LO 
pQCD formula with the CTEQ6 \cite{CTEQ6} parton distribution functions.
The nuclear modification of the parton distribution
functions for $AA$ collisions
are accounted with the EPS09 correction \cite{EPS09}
(this correction gives a small deviation of $R_{AA}$ from unity even without
the jet quenching effects).
To simulate the higher order effects, as in the PYTHIA event
generator \cite{PYTHIA}, we calculate $\alpha_{s}$
for the virtuality scale 
$cQ$ with $c=0.265$.
This gives a fairly good description 
of the $p_{T}$-dependence of the particle spectra for $pp$ collisions
(note that the normalization of hard cross sections is not important
for $R_{AA}$ at all).

We assume that the induced gluon emission stage occurs after
the DGLAP one (this approximation is reasonable
since the formation length for the leading DGLAP gluon emission
is rather small \cite{RAA08}),
and that formation of the final particle $h$ occurs
outside the QGP fireball. In this picture,
the medium-modified FF for $i\to h$ transition can be written as
\beq
D_{h/i}^{m}(Q)\approx D_{h/j}(Q_{0})
\otimes D_{j/k}^{in}\otimes D_{k/i}(Q)\,,
\label{eq:50}
\eeq
where $\otimes$ means $z$-convolution, 
$D_{k/i}$ is the DGLAP FF for $i\to k$ parton transition,
$D_{j/k}^{in}$ is the FF for $j\to k$ in-medium parton transition in the QGP
fireball, and 
$D_{h/j}$ describes the vacuum fragmentation of the parton $j$ into
the final particle $h$ outside of the QGP. 
We computed the DGLAP FFs using the PYTHIA event 
generator \cite{PYTHIA}.

For the FFs of the heavy quarks for 
$c\to D$ and $b\to B$ transitions we use 
the Peterson parametrization 
\beq
D_{M/Q}(z)\propto \frac{1}{z[1-(1/z)-\epsilon_Q/(1-z)]^{2}}
\label{eq:60}
\eeq
with $\epsilon_{c}=0.06$ and $\epsilon_{b}=0.006$.
As in \cite{RAA13}, for HFEs we write the electron $z$-distribution
for $Q\to e$ transition as a convolution
$D_{e/Q}=D_{e/M}\otimes D_{M/Q}$. We express $D_{e/M}$
for the $M\to e$ decays\footnote{Note that we ignore
  the $B\to D\to e$ process since
it gives a negligible contribution \cite{Vogt}.}
via the electron momentum spectrum $dB/dp$ in the heavy meson rest frame
\beq
D_{e/M}(z,P)=\frac{P}{4}
\int_{0}^{\infty}
dq^{2}\frac{\cosh(\phi-\theta)}
{p^{2}\cosh{\phi}}
\cdot\frac{dB}{dp}\,,
\label{eq:70}
\eeq
where $p=\sqrt{(q^{2}+m_{e}^{2})\cosh^{2}(\phi-\theta)-m_{e}^{2}}$,
$\theta=\mbox{arcsinh}(P/M)$, $\phi=\mbox{arcsinh}(zP/\sqrt{q^{2}+m_{e}^{2}})$,
$P$ is the heavy meson momentum, and $M$ is its mass.  
For $dB/dp$ in the $B/D$-meson decays
we use the CLEO data \cite{CLEO_B,CLEO_D} on the electron spectra.
We calculate the $z$-distribution of the nonprompt $D$
mesons from beauty-hadron decays, $D_{D/B}(z,P)$, using a form
similar to (\ref{eq:70}) (with replacement $m_e\to m_D$) with the $D$ meson
spectrum $dB/dp$ obtained by the BaBar Collaboration \cite{babar_B-D}.

For numerical calculation of the one gluon emission spectrum $dP/dx$
we use the representation derived in \cite{Z04_RAA}.
For the convenience of the reader formulas for calculation of
$dP/dx$ are given in Appendix.
For heavy quark masses we take $m_{c}=1.2$ GeV and $m_{b}=4.75$ GeV. 
For the gluon quasiparticle mass we take $m_{g}=400$ MeV \cite{LH}
(as in \cite{Z_hl}, for jet quenching of light hadrons). 
As in \cite{Z_hl},  we calculate the dipole cross section, which is necessary
for calculation of the imaginary
potential (\ref{eq:140}) in the Schr\"odinger equation for
calculation of $dP/dx$,
using the Debye mass from the lattice simulations of \cite{Bielefeld_Md}. 

We calculate the FFs $D_{j/k}^{in}$ for heavy quarks via the 
the one gluon spectrum $dP/dx$ in the approximation of the independent
multiple gluon emission \cite{RAA_BDMS} in the same way as
in our previous jet quenching analyses for light hadrons (see
Appendix B of \cite{RAA20} for details).
As in \cite{RAA08,Z_hl}, we treat the collisional mechanism as a
perturbation to the radiative one
by redefining the initial QGP 
temperature in calculating the radiative medium-modified FFs $D_{j/k}^{in}$.
We calculate the collisional energy loss using the 
Bjorken method \cite{Bjorken1}
with an accurate treatment of kinematics of the $2\to 2$ 
processes (the details can be found
in \cite{Z_Ecoll}).

As in \cite{Z_hl}, we take $\alpha_s(Q,T)$ in the form 
\beq
\alpha_s(Q,T) = \begin{cases}
\dfrac{4\pi}{9\log(Q^2/\Lambda_{QCD}^2)}  & \mbox{if } Q > Q_{fr}(T)\;,\\
\alpha_{s}^{fr}(T) & \mbox{if }  Q_{fr}(T)\ge Q \ge cQ_{fr}(T)\;, \\
\alpha_{s}^{fr}(T)\times(Q/cQ_{fr}(T)) & \mbox{if }  Q < cQ_{fr}(T)\;, \\
\end{cases}
\label{eq:80}
\eeq
where
$Q_{fr}=\Lambda_{QCD}\exp\left\lbrace
{2\pi}/{9\alpha_{s}^{fr}}\right\rbrace$ (in the present analysis we
take $\Lambda_{QCD}=200$ MeV), $c=0.8$.
We take $Q_{fr}=\kappa T$, and perform fit of the free parameter $\kappa$ using
data on the nuclear modification factor $R_{AA}$ for heavy ion collisions.
The form (\ref{eq:80}) is supported by the lattice results \cite{Bazavov_al1}
for the in-medium $\alpha_s$.

We use the same model of the QGP fireball as in \cite{Z_hl}
with Bjorken's 1+1D expansion of the QGP \cite{Bjorken2}
(that leads to the entropy density
$s(\tau)/s(\tau_0)=\tau_0/\tau$ with $\tau_0$ the thermalization time)
and a flat entropy profile in the
the transverse coordinates. We take $\tau_{0}=0.5$ fm.
We use a linear parametrization
$s(\tau)=s(\tau_0)\tau/\tau_0$ for $\tau<\tau_{0}$.
To fix $s(\tau_{0})$ in $AA$ collisions we use the predictions
of the Glauber wounded nucleon model \cite{KN-Glauber}
with parameters obtained in our Monte-Carlo
Glauber analyses \cite{Z_MC1,Z_MC2}
by fitting data on the charged 
hadron multiplicity pseudorapidity density $dN_{ch}/d\eta$ in $AA$
collisions
from RHIC (for 0.2 TeV Au+Au collisions) and the LHC
(for 2.76 and 5.02 TeV Pb+Pb collisions).
For the entropy/multiplicity ratio we take
$dS/dy{\Big/}dN_{ch}/d\eta\approx 7.67$ \cite{BM-entropy}.
Our Glauber model gives for the initial QGP temperature
(for the ideal gas QGP with $N_{f}=2.5$)
$T_{0}\approx 320$ MeV for central
Au+Au collisions at $\sqrt{s}=0.2$ TeV, and
$T_{0}\approx 400(420)$ MeV for central
Pb+Pb collisions at $\sqrt{s}=2.76(5.02)$ TeV
(see Fig. 1 in \cite{Z_hl}).
As in \cite{Z_hl}, we
transform the almond shaped overlap region of two colliding nuclei
into an elliptic one (of the same area), which reproduces the fireball
eccentricity $\epsilon_2$ obtained within our Monte-Carlo Glauber model.
Note that for the Monte-Carlo version of the Glauber model $\epsilon_2$
does not vanish for central collisions
(due to density fluctuations), contrary to the
optical Glauber model. This fact is practically irrelevant for $R_{AA}$,
but is important for predictions of the azimuthal anisotropy $v_2$
(see discussion in \cite{Z_hl}).

As in \cite{Z_hl}, for mQGP produced in $pp$ collisions we use the model of
an effective fireball
(that includes $pp$ collisions with all impact parameters). 
In this picture, using the data on the UE charged multiplicity density
$dN_{ch}^{ue}/d\eta$, we obtain for the radius and the initial
temperature $T_0$ of the mQGP fireball produced in $pp$ collisions
\cite{Z_hl}
\beq
R_{f}[\sqrt{s}=0.2,2.76,5.02\,\, \mbox{TeV}]
\approx[1.26,1.44,1.49]\,\,\mbox{fm}\,,
\label{eq:90}
\eeq
\beq
T_{0}[\sqrt{s}=0.2,2.76,5.02\,\,\mbox{TeV}]
\approx[195(226),217(247),226(256)]\,\,\mbox{MeV}\,.
\label{eq:100}
\eeq
In (\ref{eq:100}) we present $T_0$ for the ideal gas entropy and for the
lattice entropy \cite{t-lat}
(numbers in brackets).

For $pp$ collisions we calculate the medium-modified hard cross sections
in the same way as for $AA$ collisions. We calculate the $L$-distribution
of the jet path lengths in the mQGP fireball using the distribution
of the jet production points for the MIT bag model
quark density (assuming the same density for quarks and gluons).

\begin{figure} 
\begin{center}
\includegraphics[width=13cm]{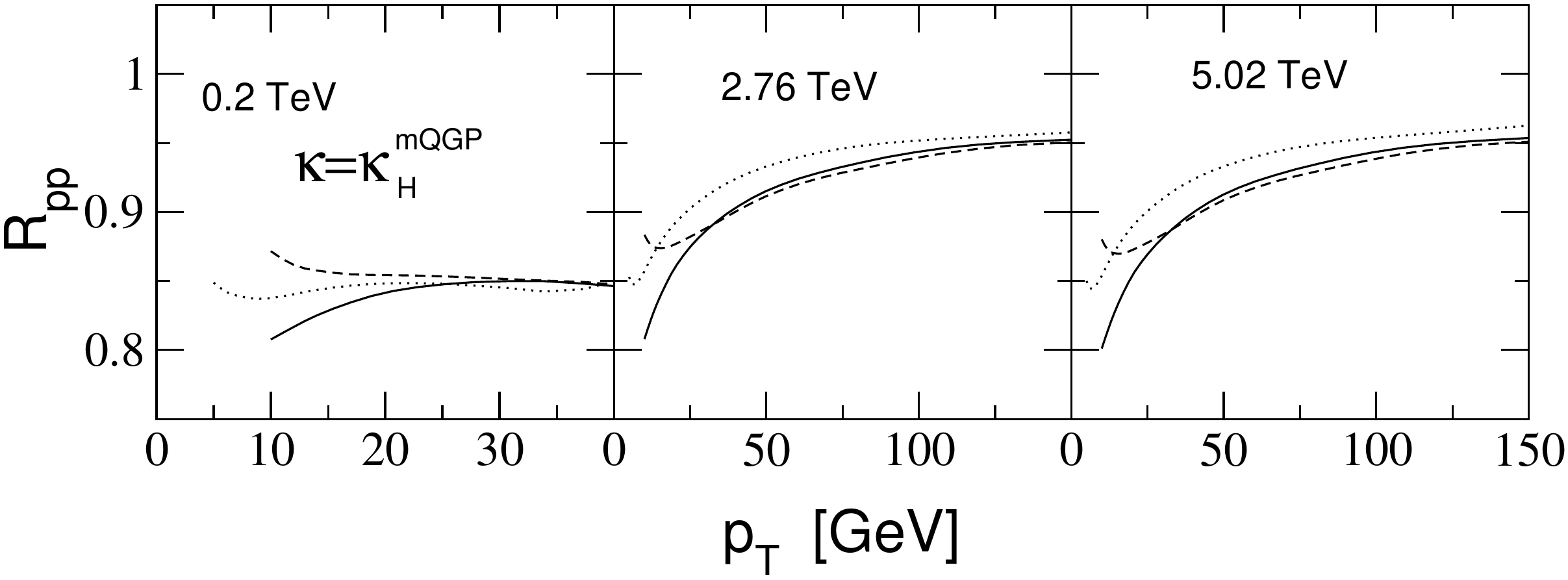}  
\includegraphics[width=13cm]{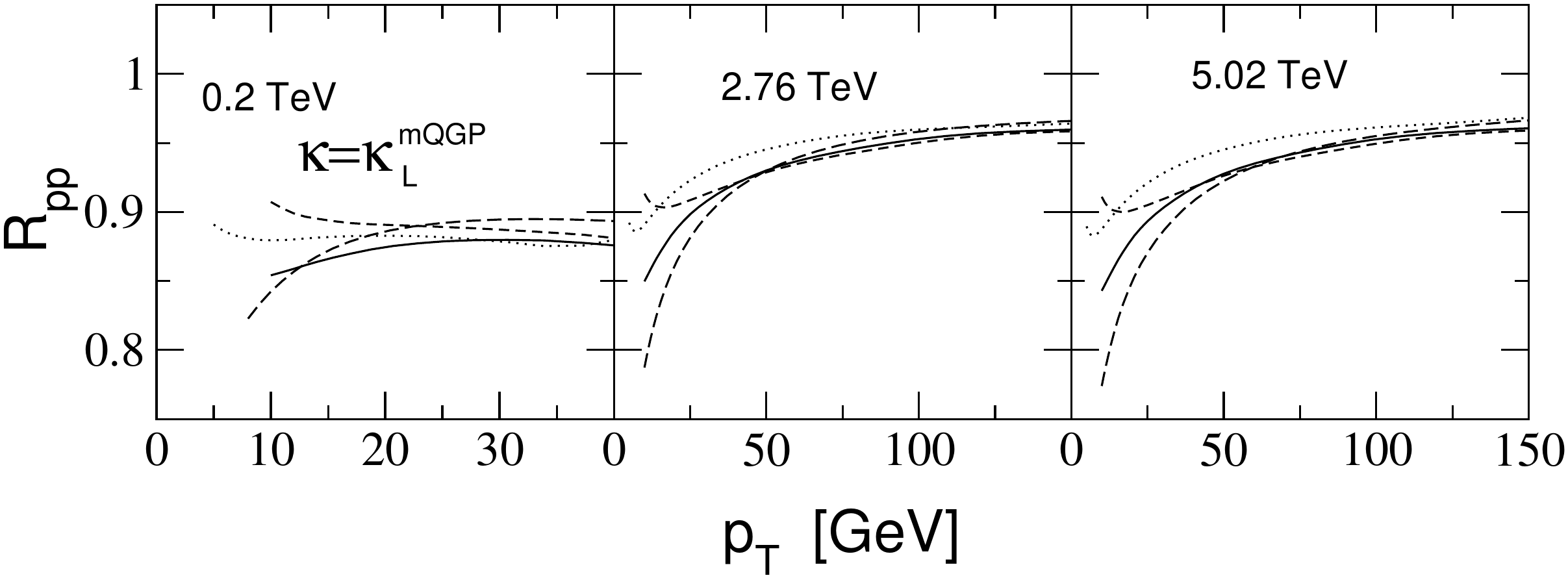}  
\end{center}
\caption[.]{
  $R_{pp}$ of $D$ mesons (solid), $B$ mesons (dashed), and HFEs (dotted)
 for $0.2$, $2.76$, and $5.02$ TeV
 $pp$ collisions. In the upper(lower) panels the curves are for
 $\kappa=\kappa_{H}^{mQGP}(\kappa_{L}^{mQGP})$.
 In the lower panels we also plot $R_{pp}$ for charged
 hadrons (long-dashed).
}
\end{figure}
\begin{figure} 
\begin{center}
\includegraphics[width=11cm]{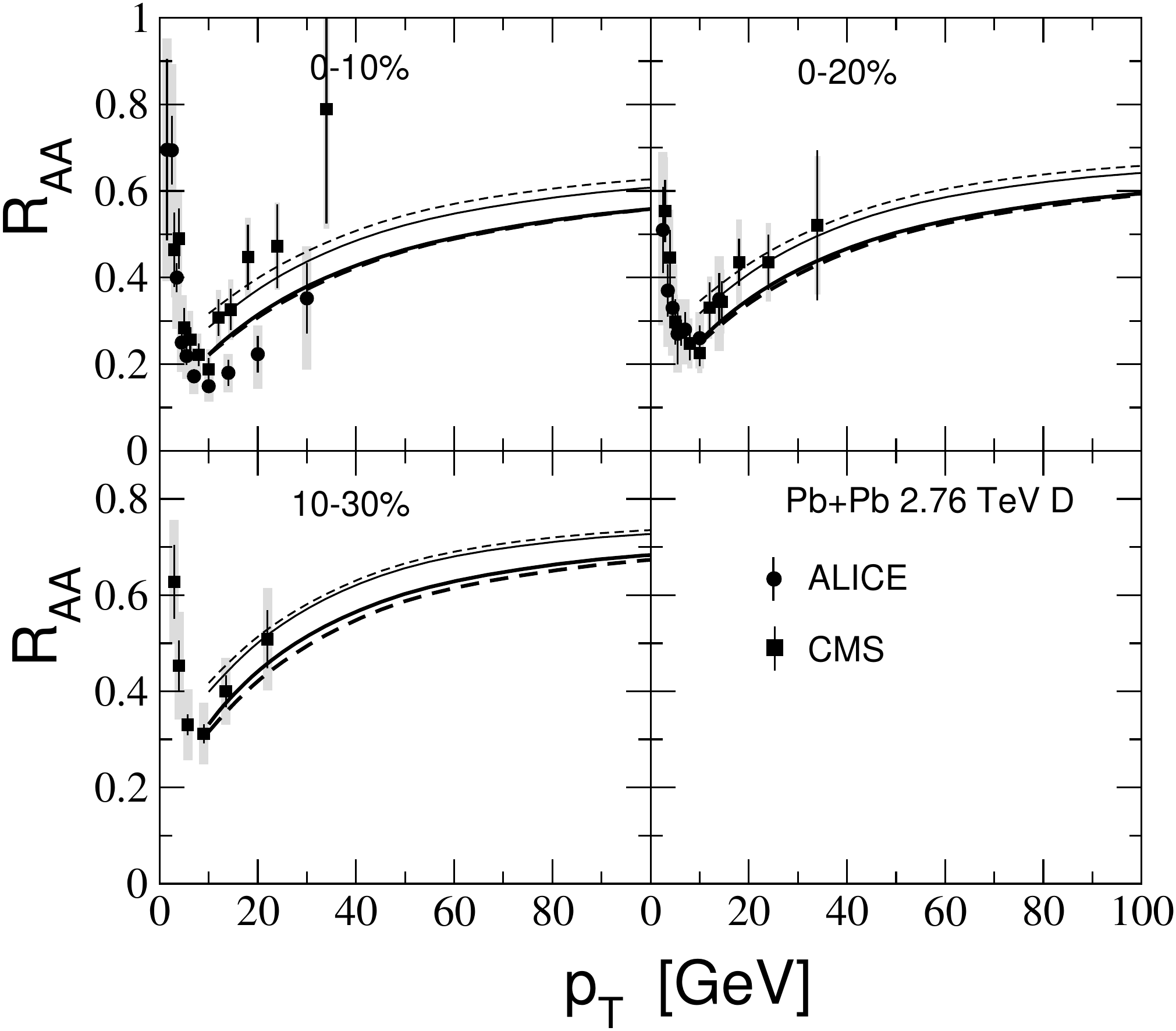}  
\end{center}
\caption[.]{
$R_{AA}$ of $D$ mesons for $2.76$ TeV Pb+Pb collisions
  from our calculations for scenarios with (solid) and without (dashed)
  mQGP formation in $pp$ collisions for
  the optimal parameters $\kappa_H^{mQGP}$($\kappa_H$)
(thick lines) and $\kappa_L^{mQGP}$($\kappa_L$) (thin lines).
  Data points are from ALICE \cite{ALICE_rD276}
and CMS \cite{CMS_rD276}.}
\end{figure}
\begin{figure} 
\begin{center}
\includegraphics[width=11cm]{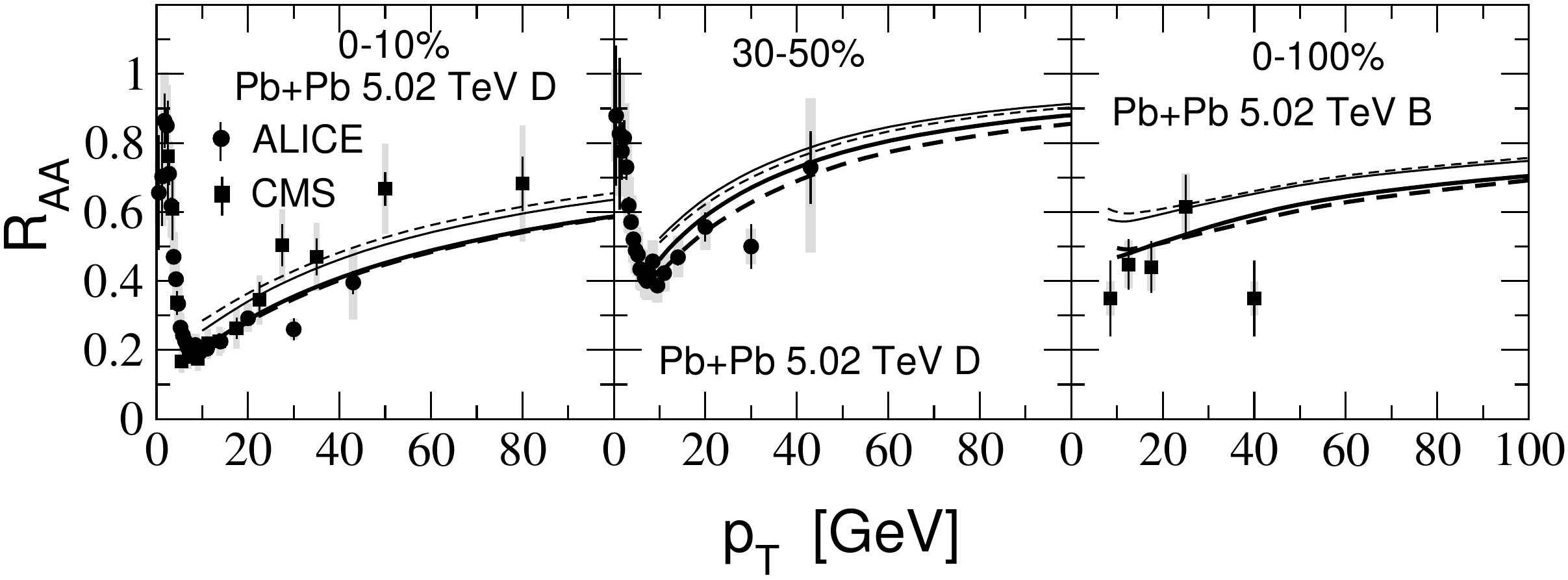}  
\end{center}
\caption[.]
{
          $R_{AA}$ of $D$ mesons (left and middle plots)
          and $B$ mesons (right plot) for $5.02$ TeV Pb+Pb collisions.
Curves are as in Fig.~2.          
  Data points for $D$ mesons are from ALICE \cite{ALICE_rD502}
and CMS \cite{CMS_rD502}, and for $B$ mesons from CMS \cite{CMS_rB502}.}
\end{figure}
\begin{figure} 
\begin{center}
\includegraphics[width=11cm]{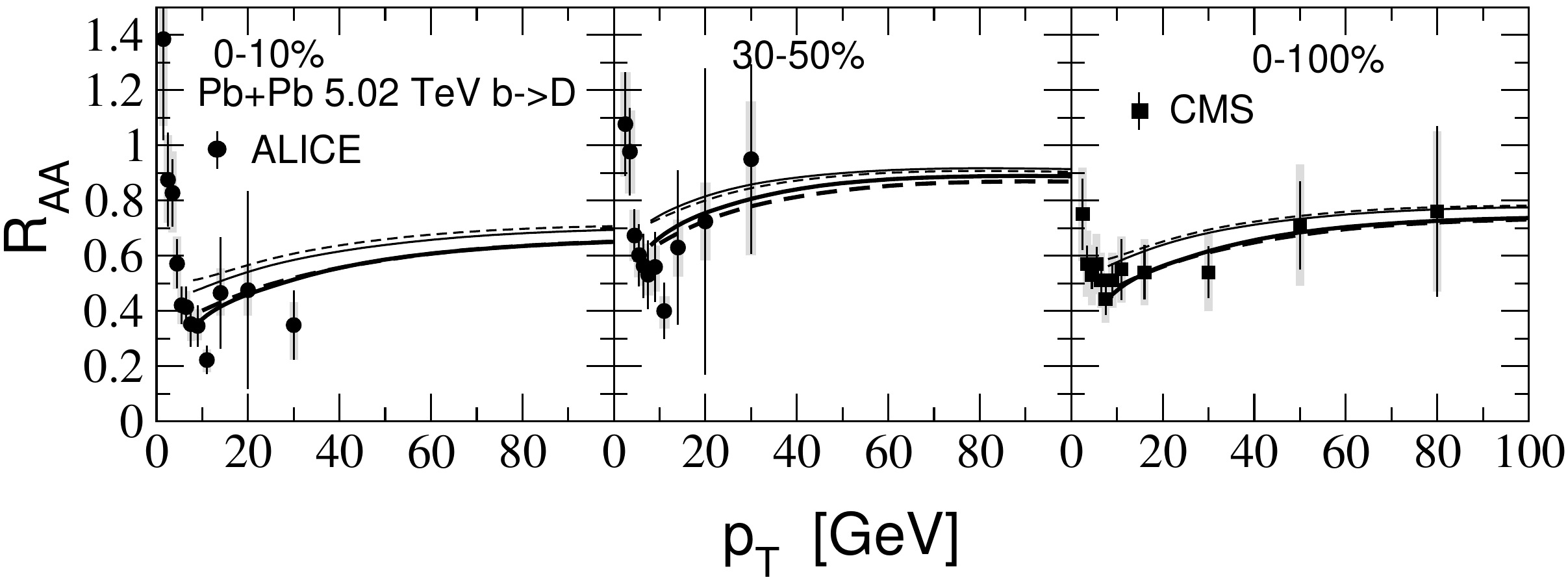}  
\end{center}
\caption[.]
        {
  $R_{AA}$ of nonprompt $D$ mesons from $B\to D^0$ decays.
  Curves are as in Fig.~2.
Data points are from ALICE \cite{ALICE_rBD502}
and CMS \cite{CMS_rBD502}.
}
\end{figure}
\begin{figure} 
\begin{center}
\includegraphics[width=10cm]{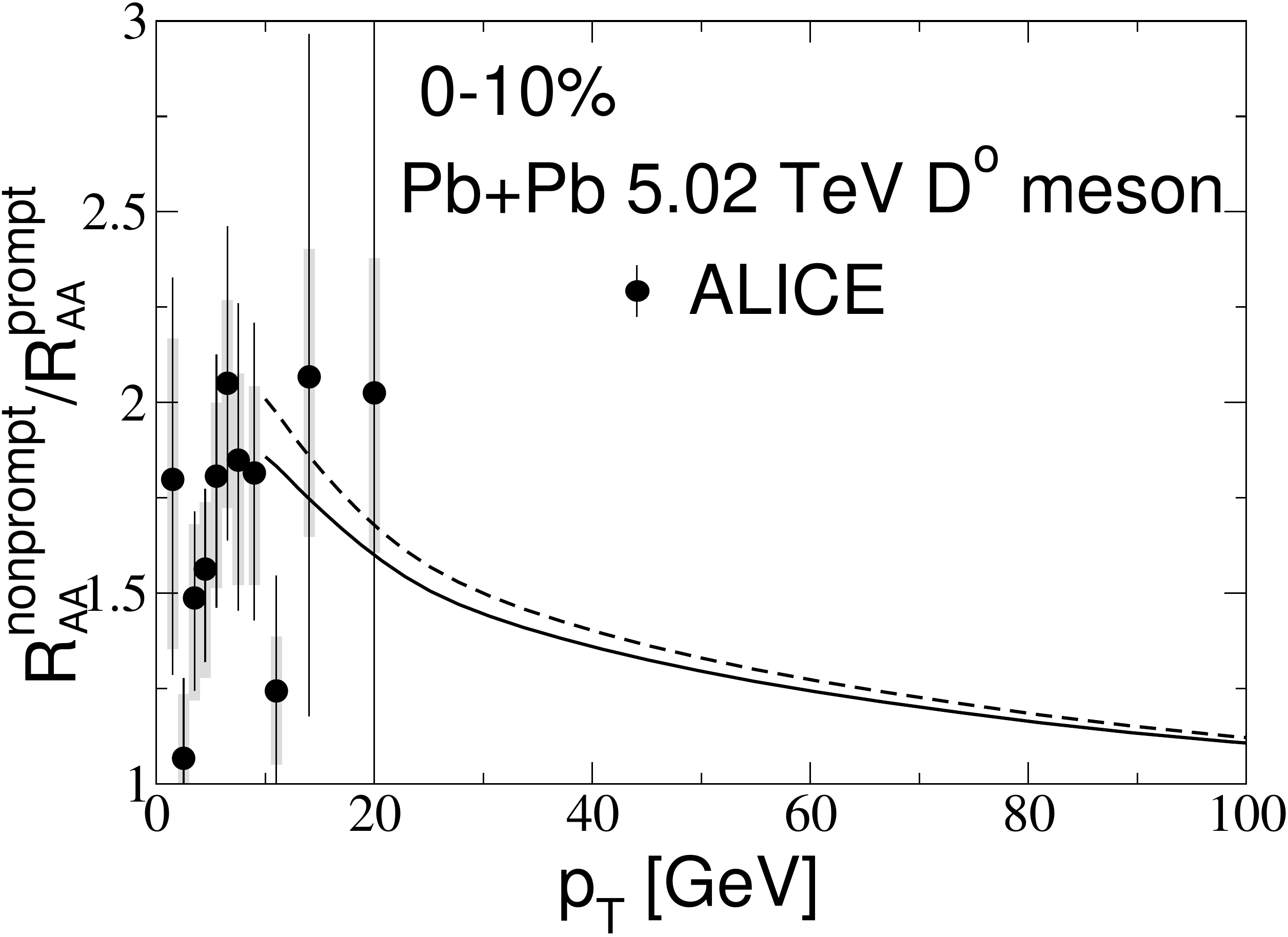}  
\end{center}
\caption[.]
        {
          Nonprompt to prompt $D^0$-meson $R_{AA}$ ratio vs $p_T$
          in the 0--10\% central 5.02 TeV Pb+Pb collisions
  from our calculations for scenarios with (solid) and without (dashed)
  mQGP formation in $pp$ collisions for
  the optimal parameters $\kappa_H^{mQGP}$ and $\kappa_H$.
          Data points are from ALICE \cite{ALICE_rBD502}.
}
\end{figure}
\begin{figure} 
\begin{center}
\includegraphics[width=11cm]{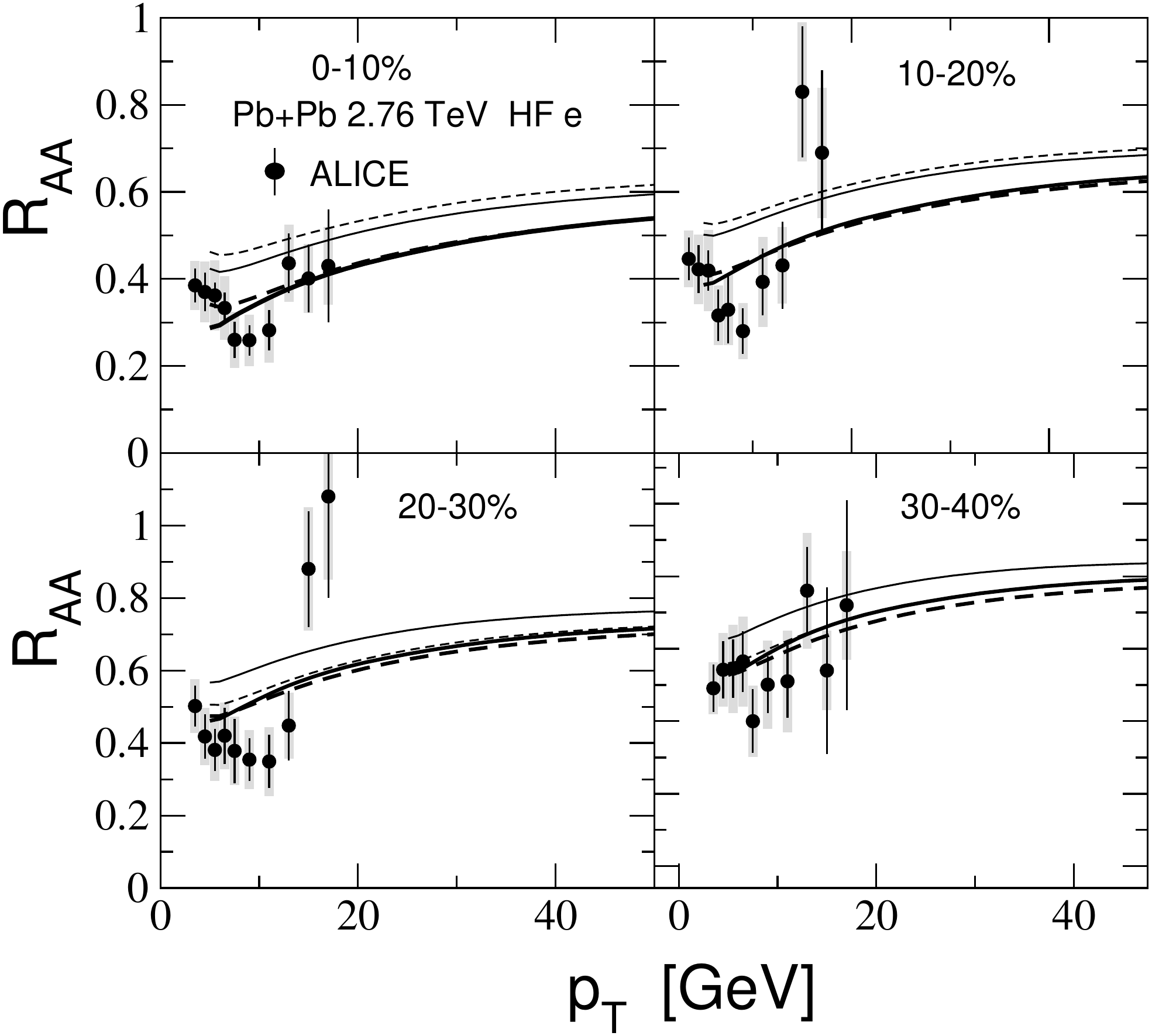}  
\end{center}
\caption[.]
        {
          Same as in Fig.~2 for HFEs at $\sqrt{s}=2.76$ TeV.
          Data points are from ALICE \cite{ALICE_rE276}.
}
\end{figure}
\begin{figure} 
\begin{center}
\includegraphics[width=11cm]{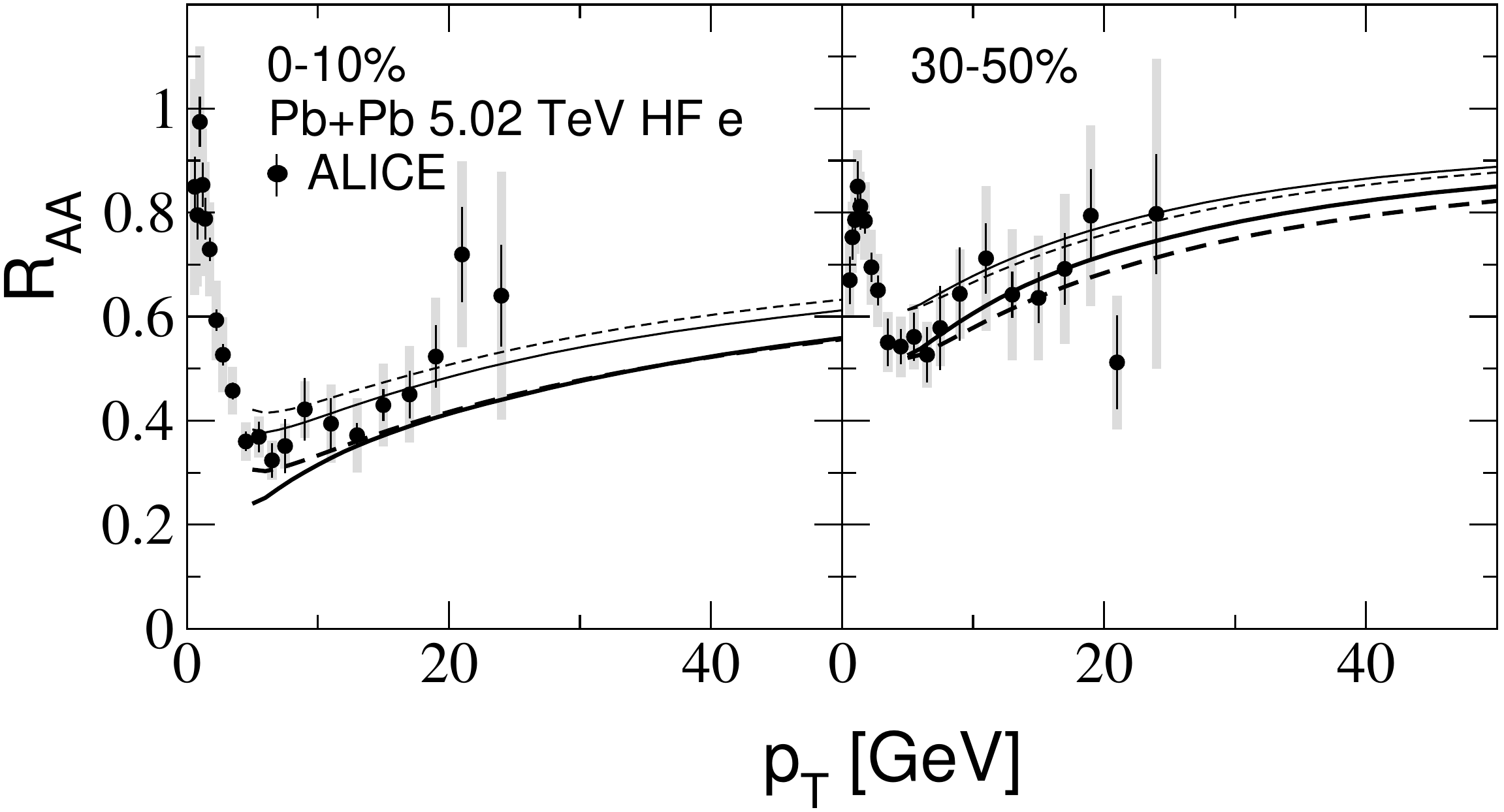}  
\end{center}
\caption[.]
{
          Same as in Fig.~6 for $\sqrt{s}=5.02$ TeV.
          Data points are from ALICE \cite{ALICE_rE502}.
}
\end{figure}
\begin{figure} 
\begin{center}
\includegraphics[width=11cm]{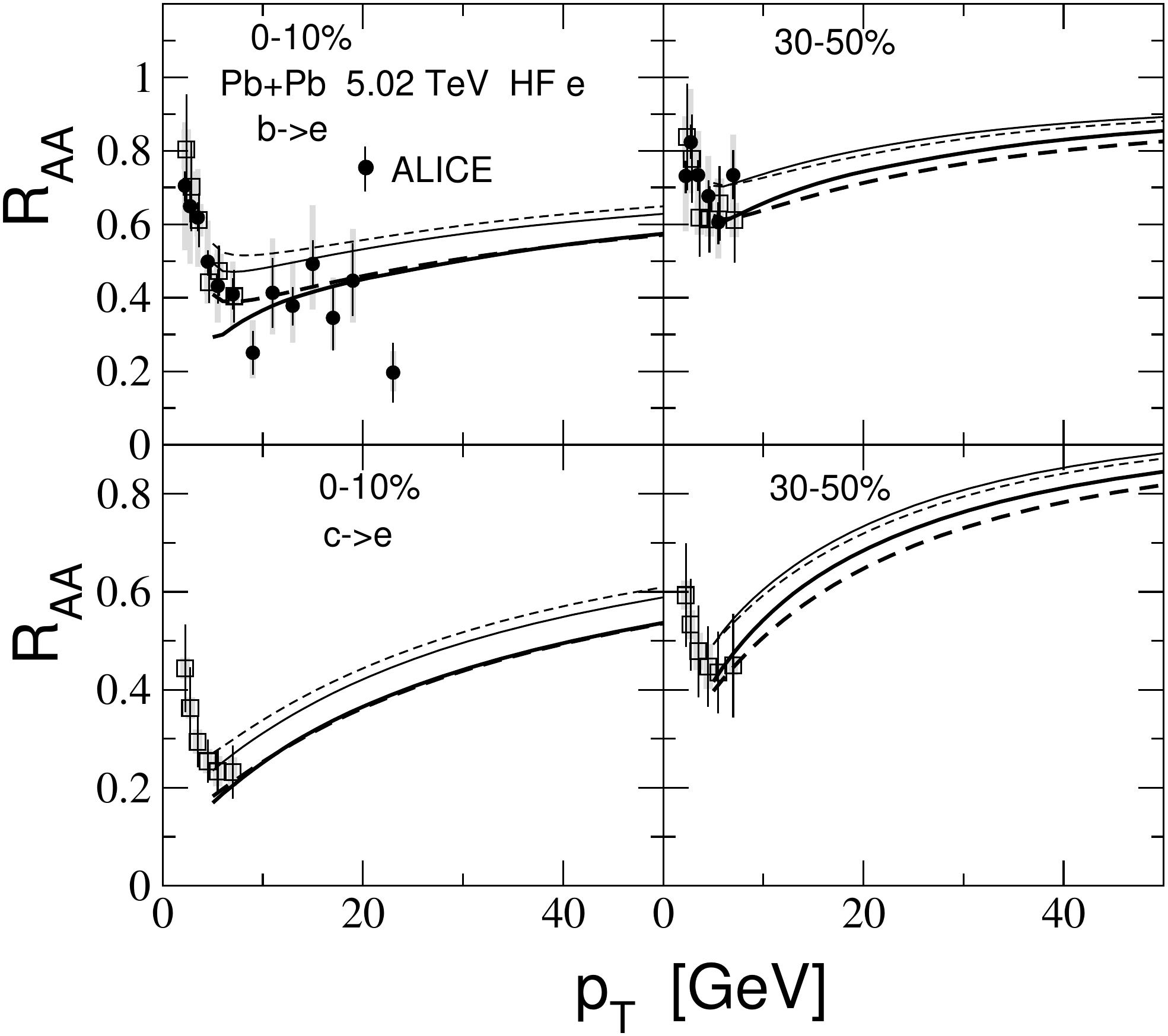}  
\end{center}
\caption[.]
        {
  $R_{AA}$ of electrons from bottom (upper part) and charm (lower part) quark
decays in  Pb+Pb collisions at $\sqrt{s}=5.02$ TeV. 
Curves are as in Fig.~2.
Data points are from ALICE \cite{ALICE_rEbc502} (circles)
and from the analysis by D. Li {\it et al.} \cite{Li_rEbc502}
(squares) within
a data-driven method of charm and bottom quark isolation.
}
\end{figure}
\begin{figure} 
\begin{center}
\includegraphics[width=11cm]{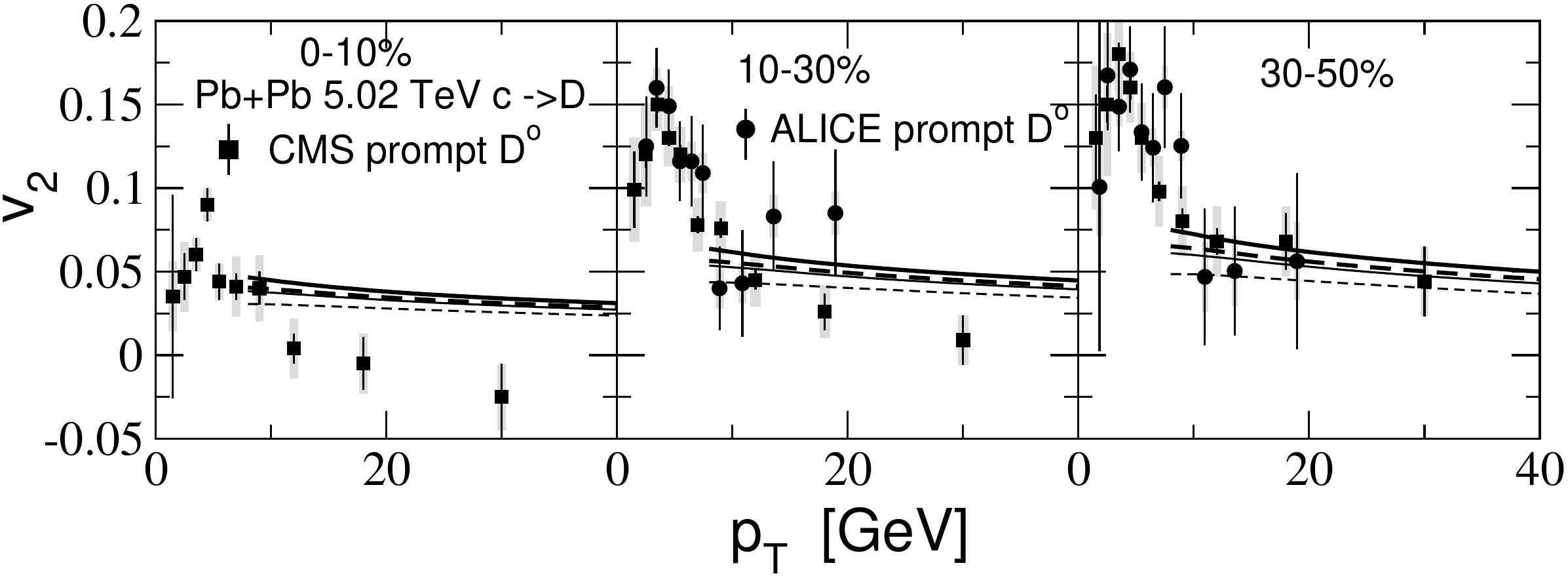}  
\includegraphics[width=11cm]{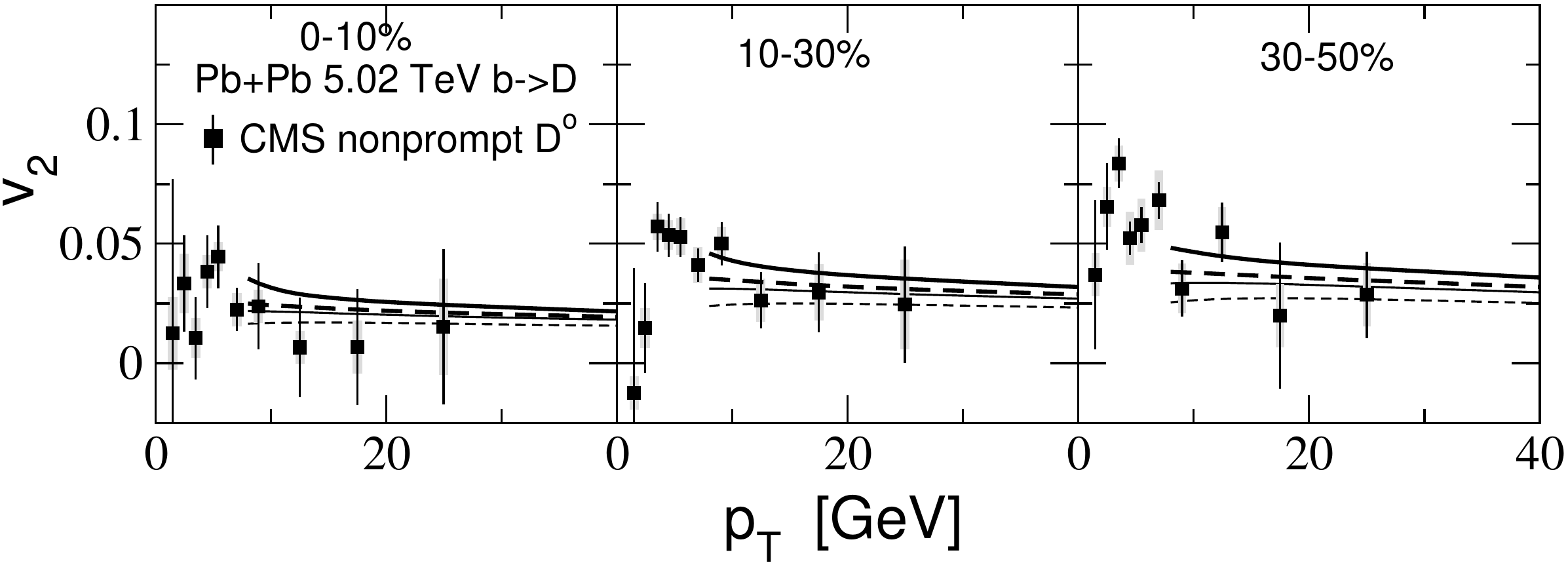}  
\end{center}
\caption[.]
        {
  $v_2$ for prompt (upper) and nonprompt (lower) $D$ mesons
in 5.02 TeV Pb+Pb collisions. 
Curves are as in Fig. 2. 
Data points are from
ALICE \cite{ALICE_v2CD502} and CMS \cite{CMS_v2CD502,CMS_v2BD502}.
}
\end{figure}
\begin{figure} 
\begin{center}
\includegraphics[width=10cm]{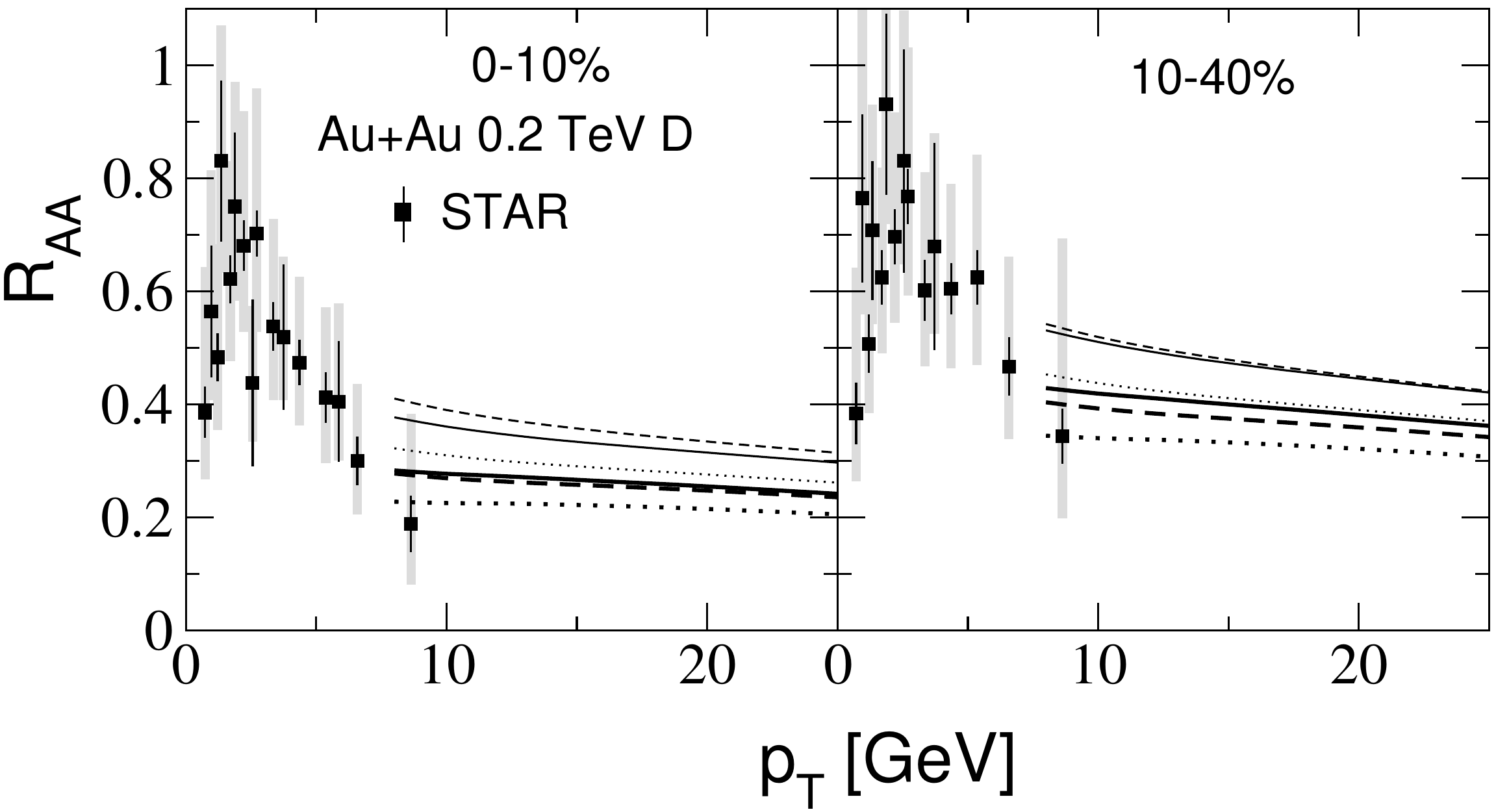}  
\end{center}
\caption[.]
        {
$R_{AA}$ of $D^0$ mesons in Au+Au collisions at $\sqrt{s}=200$ 
GeV for 0--10\% (left) and 10--40\% (right) centrality
bins.  
The solid and dashed lines are same as in Fig.~2.
The dotted lines show results for scenario without mQGP
formation in $pp$ collisions obtained with parameters
$\kappa_L^{mQGP}$ (thin lines) and $\kappa_H^{mQGP}$ (thick lines).
Data points are from STAR \cite{STAR_rD02}.
}
\end{figure}
\begin{figure} 
\begin{center}
\includegraphics[width=10cm]{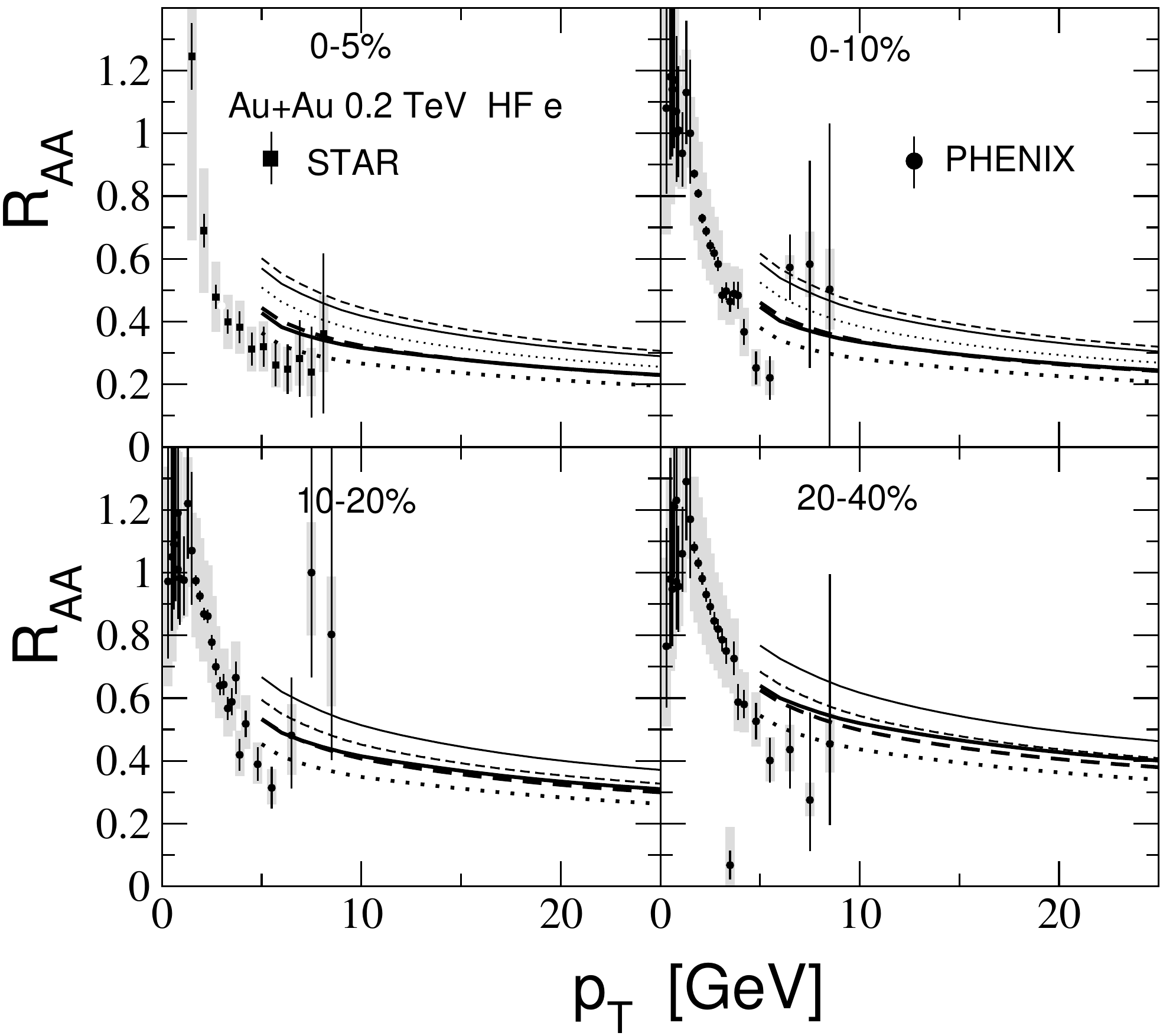}  
\end{center}
\caption[.]
{
  Same as in Fig. 10 for HFEs.
  Data points are from STAR \cite{STAR_e}
  and PHENIX \cite{PHENIX2_e}.
}
\end{figure}
\begin{figure} 
\begin{center}
\includegraphics[width=11cm]{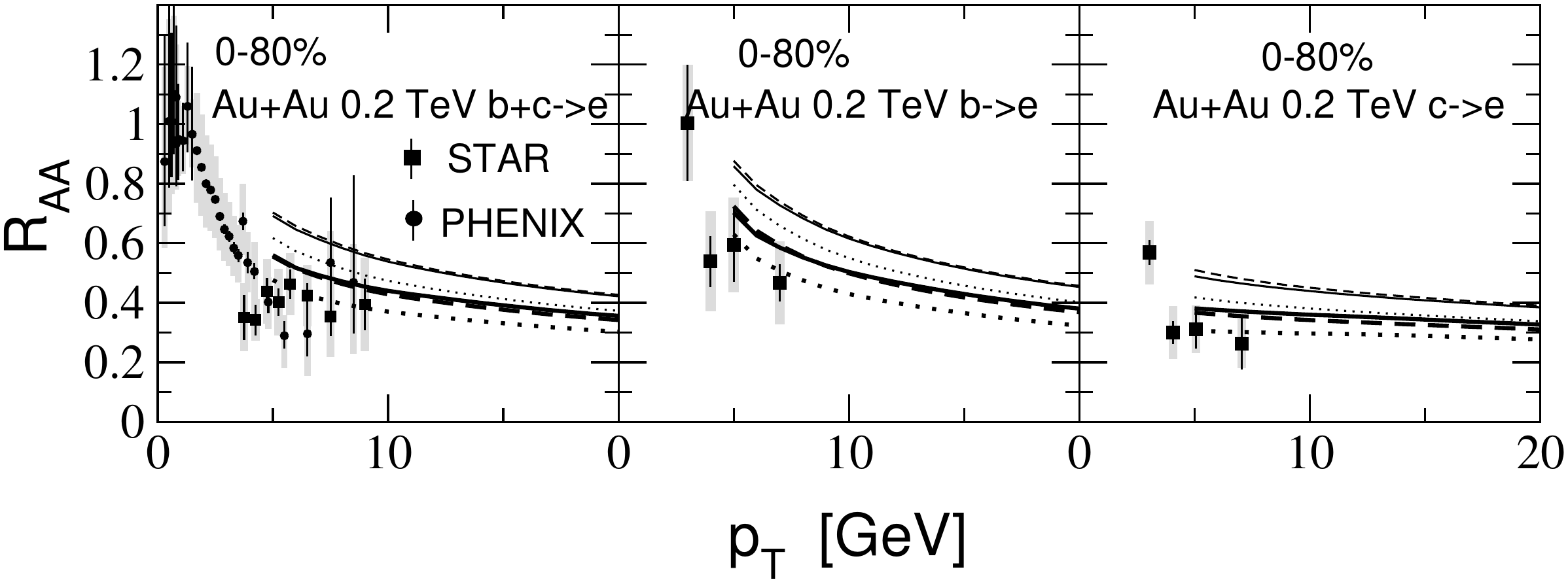}  
\end{center}
\caption[.]
        {
The HFE $R_{AA}$ in Au+Au collisions at
          $\sqrt{s}=0.2$ TeV for minimum-bias centrality 0--80\% interval.
(left) The inclusive $b+c\to e$ $R_{AA}$. (middle) $R_{AA}$ for
bottom-decay electrons. (right)          
$R_{AA}$ for charm-decay electrons. Curves are as in Fig. 10. 
Data points are from PHENIX \cite{PHENIX2_e} and 
STAR \cite{STAR_ebc}.
}
\end{figure}

\section{Comparison with experimental data}
In this section we compare the model predictions
with data for the nuclear modification factor $R_{AA}$ and
the azimuthal anisotropy $v_2$ for heavy mesons and HFEs.
We present results for two sets of the optimal values of
the free parameter $\kappa$ in the parametrization  (\ref{eq:80})
of $\alpha_s$. The first set (for the versions
with and without mQGP formation in $pp$ collisions)
of the optimal values of $\kappa$ have been obtained 
by the $\chi^2$ fitting data from the LHC
on $R_{AA}$ for $D$ mesons
\cite{ALICE_rD276,CMS_rD276,ALICE_rD502,CMS_rD502} and HFEs
\cite{ALICE_rE276,ALICE_rE502} 
    in 2.76 and 5.02 TeV Pb+Pb collisions
with centralities $\lsim 50$\%.
We used data points with $p_T\gsim 10$ GeV for $D$ mesons,
and $p_T\gsim 5$ GeV for HFEs \footnote{For HFEs we use a smaller lower
  limit of $p_T$ since for HFEs, due to the presence of the
  additional FF $D_{e/M}$, the ratio of the typical
  transverse momentum of the original heavy quarks to the transverse
  momentum of the final detected particle for HFEs becomes bigger
  by a factor of $\sim 2$ than that for heavy mesons.}.
The fits of the heavy flavor data give
$\kappa\approx 2(1.4)$ for the versions without(with)
the mQGP formation in $pp$ collisions (in the following we denote
them as $\kappa_H(\kappa_H^{mQGP})$).
For the optimal values $\kappa_H(\kappa_H^{mQGP})$ we obtained in these fits
$\chi^2/d.p.\approx 0.69(0.71)$ ($\chi^2$ per data point).
For the second set we use
the values of $\kappa$ obtained in \cite{Z_hl} by fitting the LHC data
on $R_{AA}$ for charged hadrons
for 2.76  and 5.02 TeV Pb+Pb collisions, 
and 5.44 TeV Xe+Xe collisions. These fits
give $\kappa\approx 3.4(2.5)$
for the scenarios without(with) the mQGP production in $pp$
collisions (in the following we denote
them as $\kappa_L(\kappa_L^{mQGP})$).
For the optimal values $\kappa_L(\kappa_L^{mQGP})$, obtained by fitting
$R_{AA}$ for charged hadrons, we have for the heavy flavor data
the values $\chi^2/d.p.\approx 1.95(1.45)$, that show that  
$\kappa_L(\kappa_L^{mQGP})$ also lead to reasonable agreement with the
heavy flavor experimental data.

In Fig.~1 we show the results for $R_{pp}$ obtained for the
optimal value $\kappa=\kappa_H^{mQGP}$ (upper panels) and
$\kappa=\kappa_{L}^{mQGP}$ (lower panels)
for $\sqrt{s}=0.2$, $\sqrt{s}=2.76$,
and $5.02$ TeV. To demonstrate
the difference between the medium effects for heavy flavors and
light hadrons, in the lower panels we also plot $R_{pp}$ for charged
hadrons. From Fig. ~1 one can see that the difference between
the heavy flavor $R_{pp}$ at the LHC energies for the $L$ and $H$ versions
of the parameter $\kappa$ becomes small at $p_T\gsim 30$ GeV. 
And at $p_T\sim 10-20$ GeV for the optimal values 
$\kappa_H(\kappa_H^{mQGP})$ heavy flavors the quantity $|R_{pp}-1|$ are larger
than those for $\kappa_L(\kappa_L^{mQGP})$
by $\sim 20-25$\%. As one can see from Fig.~1,
for the LHC energies $R_{pp}$ for heavy mesons
and light hadrons become similar at $p_T\gsim 30$ GeV.

In Fig.~2 we compare our results for $R_{AA}$ of $D$ mesons with the LHC data
from ALICE \cite{ALICE_rD276} for 2.76 TeV Pb+Pb collisions.
We show the curves 
for the scenarios with (solid) and without (dashed) mQGP formation
in $pp$ collisions for the optimal values of $\kappa$ obtained
from the LHC data on heavy flavor $R_{AA}$ (thick lines) and from $R_{AA}$ of
light hadrons (thin lines).
In Fig.~3 we show comparison of our results for $R_{AA}$ of $D$ and $B$ mesons
for 5.02 TeV Pb+Pb collisions with data from ALICE \cite{ALICE_rD502}
and CMS \cite{CMS_rD502,CMS_rB502}.
The results for $R_{AA}$ of $D$ mesons
shown in figures 2 and 3 are sensitive to the charm quark energy loss.
We also calculated $R_{AA}$ 
for $D$ mesons from $B$ hadron decays (nonprompt $D$), which
is sensitive to the bottom quark energy loss. Figure 4
shows the comparison
of our results for $R_{AA}$ of nonprompt $D$ mesons in 5.02 TeV Pb+Pb
collisions with data from ALICE \cite{ALICE_rBD502} and CMS \cite{CMS_rBD502}.
In Fig. 5 we compare results for the ratio $R_{AA}^{nonprompt}/R_{AA}^{prompt}$
with data from ALICE \cite{ALICE_rBD502}. From Figs. 3--5
one can see that the model describes reasonably the difference in the
strength of jet quenching for the prompt and nonprompt $D$ mesons
(which is sensitive to the mass dependence of the quark energy loss).
In Figs.~6 and 7 we compare our results for $R_{AA}$ of HFEs
in 2.76 and 5.02 TeV Pb+Pb collisions with data
from ALICE \cite{ALICE_rE276,ALICE_rE502}. These figures correspond
to nuclear suppression of the total HFE spectrum that includes
$c\to e$ and $b\to e$ decays. Figure 8 shows comparison
of our calculations of $R_{AA}$ for $b\to e$ and $c\to e$ channels separately
with data from ALICE \cite{ALICE_rEbc502} for the $b\to e$ channel and
with the results of analysis \cite{Li_rEbc502} within
a data-driven method of charm and beauty isolation.

From Figs.~2--8 one can see that
the difference between theoretical $R_{AA}$ for $D$ and $B$ mesons, and HFEs
for scenarios with and without the mQGP formation in $pp$ collisions
is small. One can see that both for 
$\kappa_H(\kappa_H^{mQGP})$ and $\kappa_L(\kappa_L^{mQGP})$
the results show  reasonable agreement with experimental data.
Note that the results shown in Figs. 3--5 and 8 demonstrate that the model
reproduces reasonably the relative strength  of jet quenching for charm
and bottom quarks (i.e. the model reproduces reasonably the quark mass
effects).

In Fig.~9 we compare our results for $v_2$ of prompt and nonprompt
$D$ mesons in 5.02 TeV Pb+Pb collisions to data from
ALICE \cite{ALICE_v2CD502} and CMS \cite{CMS_v2CD502,CMS_v2BD502}.
Unfortunately, experimental errors are too large to make a conclusive
statement on agreement with the data.
From Fig. 9 one sees that the relative effect of the mQGP formation
in $pp$ collisions on the theoretical predictions for $v_2$ is more pronounced
than for $R_{AA}$. This occurs because the scenario with the mQGP
formation in $pp$ collisions requires somewhat stronger
jet quenching for particle spectra than that without
the mQGP formation in $pp$ collisions (to compensate
the effect of the $1/R_{pp}$ factor on $R_{AA}$).
As a consequence, for scenario with the mQGP formation we have
a larger azimuthal anisotropy $v_2$, which is not affected
by the $1/R_{pp}$ factor.

In Fig.~10 we compare our results for $R_{AA}$ of $D$ mesons in
0.2 TeV Au+Au collisions to data from STAR \cite{STAR_rD02}.
In this figure, in addition to the scenarios with and without
mQGP formation in $pp$ collisions, we also present predictions
for an intermediate scenario, in which the mQGP
production in $pp$ collisions occurs only at the LHC energies.
In this scenario $R_{AA}$ for $0.2$ TeV Au+Au collisions
should be calculated without $1/R_{pp}$ factor for the optimal
$\kappa$ fixed from the LHC data on $R_{AA}$ for
the scenario with the mQGP production in $pp$ collisions
(i.e., for $\kappa=\kappa_{L,H}^{mQGP}$).
Unfortunately, the STAR data \cite{STAR_rD02} are restricted to rather low
transverse momenta, where the applicability of our model may be questionable.
From Fig.~10 one can see that, for the maximal transverse momentum ($p_T\sim 8$
GeV) in the STAR data, our results, within errors, are consistent
with the experimental data. We get a somewhat better agreement with the data
for the intermediate scenario with the mQGP formation in $pp$ collisions
only at the LHC energies.

In Fig.~11 we compare our predictions for $R_{AA}$ of the HFEs
for $b+c\to e$ decays in
0.2 TeV Au+Au collisions to RHIC data from
STAR \cite{STAR_e} and PHENIX \cite{PHENIX2_e}.
Figure 12 shows comparison to data on $R_{AA}$ from STAR \cite{STAR_ebc}
for the total ($b+c\to e$) electron spectrum and separately
for $b\to e$ and $c\to e$ channels.
From Figs. 11 and 12 one can see that for $R_{AA}$ of HFEs, as in the case of
the results for $R_{AA}$ of $D$ mesons shown in Fig. 10, the agreement with
the experimental data becomes somewhat better for the intermediate
scenario with the
mQGP formation in $pp$ collisions only at the LHC energies.
However, a definite conclusion cannot be drawn given large experimental errors
and a very restricted $p_{T}$ range ($p_T\lsim 8$ GeV) of the data.

Thus, from Figs. 2--12 we can conclude that altogether our theoretical
results for scenarios with and without mQGP formation in $pp$ collisions
agree reasonably with experimental data on jet quenching
for heavy flavors. However,
our fits to heavy flavor $R_{AA}$ give smaller values of $\kappa$ than those
for light hadrons, i.e. heavy flavor jet quenching data require somewhat
bigger $\alpha_s$ than data on jet quenching for light hadrons.
This inconsistency could be due to the approximations used in calculations
of $R_{AA}$ from the one gluon emission spectrum.
One of the possible reasons is the use of the approximation
of independent gluon emission \cite{RAA_BDMS} for the
multiple gluon radiation. One can expect that this approximation
becomes less reliable for gluons. Since at the LHC energies
the gluon contribution to the high-$p_T$ light hadron spectrum is large,
it is clear that the different levels of inaccuracy of this approximation
for quarks and gluons can lead to
an inconsistency in the optimal values of $\kappa$ fitted to data on $R_{AA}$
for heavy flavors and light hadrons. 
Also, some inconsistency between the optimal $\kappa$ for
heavy flavors and light hadrons, may arise due to the
approximation of a flat fireball density, because this approximation
may somewhat overestimate the effect of the boundary gluon emission
which becomes stronger for gluons.

\section{Summary}
In this paper we presented results of a global analysis of experimental
data on jet quenching for heavy flavors (for $D$, $B$ mesons, and HFEs) within
the LCPI \cite{LCPI1} approach to induced gluon emission
for scenarios with and without mQGP formation in $pp$ collisions.
The present analysis extends to heavy flavors our previous
jet quenching analysis for light hadrons \cite{Z_hl}.
As in \cite{Z_hl}, we perform calculations for a temperature dependent
running coupling $\alpha_s(Q,T)$, which has a plateau
around $Q\sim Q_{fr}= \kappa T$. This parametrization is motivated by the
lattice calculation \cite{Bazavov_al1} of the in-medium QCD coupling in the QGP.
We performed calculations for two sets of the optimal values of
the parameter $\kappa$. For the first set we use $\kappa$ fitted to the
LHC data on the heavy flavor $R_{AA}$ in $2.76$ and $5.02$ TeV Pb+Pb
collisions, and for the second set we use $\kappa$ fitted
to the LHC data on $R_{AA}$ of light hadrons in $2.76$ and $5.02$ TeV Pb+Pb, and $5.44$ TeV Xe+Xe collisions.
We find that fits to heavy flavor $R_{AA}$ give smaller values
of $\kappa$ than those
for light hadrons, i.e. heavy flavor jet quenching data require somewhat
bigger $\alpha_s$ than data on jet quenching for light hadrons.
But the difference in the quality of agreement of the theoretical results
with experimental data for heavy flavors for two sets of $\kappa$
is not significant.

We find that the theoretical predictions for the nuclear modification
factor $R_{AA}$ for
heavy flavors at the LHC energies for scenarios with and without mQGP formation
in $pp$ collisions are very similar, but
the effect of the mQGP formation in $pp$ collisions on predictions
for azimuthal asymmetry $v_2$ is more pronounced.
The results for $R_{AA}$ and $v_2$ agree reasonably
with the LHC data both for $\kappa$ fitted to $R_{AA}$ for heavy flavor
and $R_{AA}$ for light hadrons.
The model reproduces reasonably the experimental relative strength
of jet quenching for charm and bottom quarks (i.e. it reproduces
reasonably the quark mass effects).

Note that, similarly to results of our analysis of jet quenching for light
hadrons \cite{Z_hl}, from comparison with the RHIC data on $R_{AA}$
of $D$ mesons and of HFEs,  we find
that the agreement with data at the RHIC energies
becomes somewhat better for the
intermediate scenario, in which the mQGP formation in $pp$ collisions occurs
only at the LHC energies. This is also supported by our
analysis \cite{Z_Ipp} of the data from ALICE \cite{ALICE_Ipp} on the
UE multiplicity dependence of the medium modification factor $I_{pp}$.

\vspace {.7 cm}
\noindent
{\large\bf Acknowledgements}

\noindent
This work is supported by the State program  0033-2019-0005.

\section*{Appendix}
In this appendix we give, for the convenience of the reader,
formulas for calculation of the gluon emission $x$-spectrum $dP/dx$.
We use the representation of the induced gluon spectrum 
obtained in Ref.~\cite{Z04_RAA} with the prescription of \cite{RAA20T}
for incorporating the $T$-dependent running $\alpha_s$.
For a fast quark with momentum along the $z$-axis produced
at $z=0$ in the matter of thickness $L$,  
$dP/dx$ has the form
\beq
\frac{d P}{d
x}=
\int\limits_{0}^{L}\! d z\,
n(z)
\frac{d
\sigma_{eff}^{BH}(x,z)}{dx}\,,
\label{eq:110}
\eeq
where $n(z)$ is the medium number density, $d\sigma^{BH}_{eff}/dx$ 
is an effective Bethe-Heitler cross section
for $q\to gq$ process,
given by
\bea
\frac{d
\sigma_{eff}^{BH}(x,z)}{dx}=-\frac{P_{q}^{g}(x)}
{\pi M}\mbox{Im}
\int\limits_{0}^{z} d\xi
\sqrt{\alpha_s(Q(\xi),T(z-\xi))\alpha_s(Q(\xi),T(z+\xi))}
\nonumber\\
\times
\left.
\exp{\left(-i\frac{\xi}{L_f}\right)}
\frac{\partial }{\partial \rho}\left(\frac{F(\xi,\rho)}{\sqrt{\rho}}\right)
\right|_{\rho=0}\,\,.
\label{eq:120}
\eea
Here 
$P_{q}^{g}(x)=(4/3)[1+(1-x)^2]/x$
is the ordinary pQCD $q\to g$ splitting function,
$M=E_qx(1-x)$, $L_{f}=2M/\epsilon^2$,
$\epsilon^2=m_q^2x^2+m_{g}^2(1-x)$,
$Q^{2}(\xi)=aM/\xi$ with $a\approx 1.85$ \cite{Z_Ecoll},
$F$ is the solution to the radial Schr\"odinger 
equation  
\bea
i\frac{\partial F(\xi,\rho)}{\partial \xi}=
\left[-\frac{1}{2M}\left(\frac{\partial}{\partial \rho}\right)^{2}
+v(\rho,x,z-\xi)\right.
+\left.\frac{4m^{2}-1}{8M\rho^{2}}
\right]F(\xi,\rho)\,
\label{eq:130}
\eea
with the azimuthal quantum number $m=1$, and
the boundary condition  
$F(\xi=0,\rho)=\sqrt{\rho}\sigma_{gq\bar{q}}(\rho,x,z)
\epsilon K_{1}(\epsilon \rho)$  at $\xi=0$
($K_{1}$ is the Bessel function).
The potential $v$ reads 
\beq
v(\rho,x,z)=-i\frac{n(z)\sigma_{gq\bar{q}}(\rho,x,z)}{2}\,,
\label{eq:140}
\eeq
where $\sigma_{gq\bar{q}}(\rho,x,z)$ is the three-body cross
section of interaction
of the $gq\bar{q}$ system with a medium constituent
located at $z$ ($\rho$ is the transverse
distance between $g$ and the final quark $q$). 
In the transverse plane $\bar{q}$ is located at the center of mass of
the $gq$ pair. 
The $\sigma_{gq\bar{q}}$ can be written
via the local dipole cross section
$\sigma_{q\bar{q}}(\rho,z)$ (for the color singlet $q\bar{q}$ pair)
\bea
\left.\sigma_{gq\bar{q}}(\rho,x,z)\right|_{q\to gq}=\frac{9}{8}
[\sigma_{q\bar{q}}(\rho,z)+
\sigma_{q\bar{q}}((1-x)\rho,z)]
-\frac{1}{8}\sigma_{q\bar{q}}(x\rho,z)\,.
\label{eq:150}
\eea
In the two-gluon approximation the dipole cross section reads
\beq
\sigma_{q\bar{q}}(\rho,z)=C_{T}C_{F}\int d\qb
\alpha_{s}^{2}(q,T(z))
\frac{[1-\exp(i\qb\ro)]}{[q^{2}+\mu^{2}_{D}(z)]^{2}}\,,
\label{eq:160}
\eeq
where $C_{F,T}$ are the color Casimir for the quark and thermal parton 
(quark or gluon), and $\mu_{D}(z)$ is the local Debye mass.

For the QGP fireball in $AA$ collisions the coordinate $z$ 
coincides
with the proper time $\tau$, i.e. in terms of the real fireball number
density, $n_f(\ro,\tau)$, we 
have $n(z)=n_f(\ro_j(\ro_{j0},\tau),\tau)$, where $\ro_{j0}$ is the jet
production transverse coordinate, and
$\ro_j(\ro_{j0},\tau)=\ro_{j0}+\tau {\bf{p}}_T/|{\bf{p}}_T|$ is
the jet trajectory.
We use the approximation of a uniform fireball. In this case,
inside the fireball, the function $n_f(\ro,\tau)$ does not depend on
the jet production point.
This greatly reduces the computational cost,
since one can tabulate the $L$-dependence of
the induced gluon spectrum once, and then use it
for calculations of the FFs for arbitrary jet geometry.


\end{document}